\documentclass[prb,twocolumn,showpacs]{revtex4}
\usepackage{graphicx}

\newlength{\figwidth}
\begin{document}

%
%

\title{The Compressible Ising Spin Glass:  Simulation Results}
\author{Adam H. Marshall}
\affiliation{The James Franck Institute and Department of Physics,
   University of Chicago, Chicago, Illinois 60637}
\date{\today}

\begin{abstract}
This paper reports numerical studies of a compressible version of the Ising
spin glass in two dimensions.  Compressibility is introduced by adding a
term that couples the spin-spin interactions and local lattice
deformations to the standard Edwards-Anderson model.  The relative
strength of this coupling is controlled by a single dimensionless
parameter, $\mu$.  The timescale associated with the dynamics of the
system grows exponentially as $\mu$ is increased, and the energy of the
compressible system is shifted downward by an amount proportional to $\mu$
times the square of the uncoupled energy.  This result leads to the
formulation of a simplified model that depends solely on spin variables;
analysis and numerical simulations of the simplified model predict a
critical value of the coupling strength above which the spin-glass
transition cannot exist at any temperature.
\end{abstract}


\maketitle

%
%

\section{Introduction}

Much theoretical study has been made of the nature of the spin-glass
transition.  It is now generally accepted that the three-dimesional spin
glass undergoes a second-order phase transition at finite
temperature,\cite{bhat85,ogie85,ogie85b,bind86,bhat88,fisc93,kawa96} and
the bulk of the evidence in two dimensions is consistent with a
zero-temperature phase
transition,\cite{bhat88,wang88,saul93,kita00,houd01,katz05} although
recent work suggests that the lower critical dimension for some spin-glass
models is greater than two.\cite{amor03}  The continued controversy hints
at the delicate and subtle nature of the spin-glass transition and
suggests that modifications to the underlying model, even small ones,
could have dramatic effects on the system.

Compressibility has already been shown to have a strong effect on a
variety of spin systems.  The inclusion of compressibility in the Ising
ferromagnet modifies the standard second-order transition to a first-order
transition that occurs at the Curie temperature.\cite{bean62,berg76}  
The (fully frustrated) 2-D triangular Ising anti-ferromagnet does not
undergo a phase transition; however, when compressibility is added to the
model, the characteristic frustration is relieved, and the system develops
a first-order transition to a ``striped'' phase at low
temperatures.\cite{chen86,gu__96,dhar00}  Other frustrated spin systems
are known to have their frustration relieved by the presence of
magnetoelastic couplings,\cite{becc03,sush05} and polaron effects alter
the nature of magnetic transitions in frustrated physical systems such as
manganites.\cite{sala01,laih01,adam04}

With the possibility of relieving frustration, the addition of
compressibility to spin-glass models could dramatically alter the nature
of the spin-glass phase and/or the transition thereto.  Furthermore, the
fact that all physical systems must possess some (albeit small)
spin-lattice coupling provides a physical motivation for such studies.

A previous paper introduced a particular model for the compressible spin
glass with a linear coupling between the spin-spin interactions and the
distances between neighboring particles.\cite{mars06}  The work described
there involved simulations of the compressible spin glass performed on
two-dimensional systems in which the volume was held fixed.  Results of
the direct simulations suggest a simplified model, qualitatively
equivalent to the first, that depends only upon spin degrees of freedom.
The presence of compressibility alters the preferred spin configurations
of the system, so that the transition to a low-temperature spin-glass
phase is impossible above a critical value of the coupling.  The current
paper expands on that previous work as well as provides details of the
analysis.  Presented here are results showing the exponential slowing down
of the time to reach equilibrium as the coupling increases, additional
quantitative motivation for the simplified model, and a functional form
for the entropy of the spin glass, from which thermodynamic quantities are
predicted.  Finally, a phase diagram illustrates an approximate boundary
separating critical behavior from the region where the spin-glass
transition cannot exist.

The structure of this paper is as follows:  Section~\ref{sec_model}
describes the Hamiltonian of the compressible spin glass and defines the
important tuning parameters.  The details of the computer simulations are
discussed in Sec.~\ref{sec_details}, and the results of the simulations
are presented in Sec.~\ref{sec_results}.  A simplified model is introduced
in Sec.~\ref{sec_simplified}, along with results of numeric simulations
and analytic investigations performed on the simplified model.
Section~\ref{sec_conclusion} contains the primary conclusions and some
additional points of discussion.

\section{The Model}
\label{sec_model}

The Hamiltonian for the compressible Ising spin glass is\cite{mars06}
\begin{equation}
   \mathcal{H} = - \sum_{\langle i,j \rangle} J_{ij} S_i S_j
      + \alpha \sum_{\langle i,j \rangle} J_{ij} S_i S_j
            \left( r_{ij} - r_0 \right)
      + U_\mathrm{lattice}  \, .
\label{eqn_csg_ham}
\end{equation}

The first term is the standard Edwards-Anderson spin-glass
Hamiltonian,\cite{edwa75} with the sum performed over pairs of nearest
neighbors.  The spins $S_i$ are dynamic variables which may take the
values $+1$ or $-1$.  The interactions $J_{ij}$ are chosen randomly from
$\{\pm J\}$ with equal probability and are then held fixed; this
collection of interactions represents a single realization of the quenched
disorder central to the nature of the spin glass.

The coupling between the spin interactions and the lattice distortions is
contained within the second term of Eq.~(\ref{eqn_csg_ham}), where the
coupling is considered to linear order with proportionality constant
$\alpha$.  This constant multiplies the change in bond length:  $r_{ij}$
represents the Euclidean distance between particles $i$ and $j$, and $r_0$
is the natural spacing of nearest neighbors on the lattice.  This term
allows the system to lower the total energy by displacing the particles
from their regular lattice positions.  Spins with satisfied interactions
(i.e., those with $J_{ij} S_i S_j = +1$) will tend to move closer together
in order to strengthen the effect; similarly, unsatisfied bonds will tend
to lengthen as the particles move farther apart to diminish the negative
effect of their interaction on the total energy.  The inability of all of
the bonds in the system to distort simultaneously in the ideal fashion is
the mechanism by which the degeneracy of configurations with equal
spin-spin energy is broken.

The final term, $U_\mathrm{lattice}$, stabilizes the lattice by providing
a restoring force to counteract the displacements generated by the
spin-lattice interactions.  The stabilization is obtained by connecting
harmonic springs between nearest neighbors and between next-nearest
neighbors (along the diagonals of the square lattice).  Each spring has as
its unstretched length the natural spacing of the vertices, so that
$U_\mathrm{lattice}$ is zero in the absence of spin-lattice coupling
when the particles are not displaced.

Two important parameters can be formed.  The first arises due to force
balance between the last two terms in Eq.~(\ref{eqn_csg_ham}):
\begin{equation}
   \overline{\delta} \equiv \frac{J \alpha}{k}  \, ,
\label{eqn_delta_def}
\end{equation}
This has dimensions of length and represents the scale of the typical
displacements of the particles from their uncoupled locations on the
square lattice.  The second parameter is
\begin{equation}
   \mu \equiv \frac{J \alpha^2}{k}  \, ,
\end{equation}
which is dimensionless; it represents the strength of the spin-lattice
coupling, relative to the spin-spin interaction.  When $\mu = 0$, there is
no spin-lattice coupling, lattice distortions are not energetically
favorable, and the model reduces to the standard Edwards-Anderson spin
glass.  The interaction strength $J$ serves merely to set the energy scale
for the model.  In this work, $J$ is set to unity, while $\alpha$ and $k$
are chosen so that $\overline{\delta}$ and $\mu$ take the desired values.

\section{Simulation Details}
\label{sec_details}

All simulations were run on square lattices of linear dimension $L$ with
periodic boundary conditions in two dimensions; the size of the system
was held constant in each direction, fixing the total volume.  The control
parameters were adjusted so that $\overline{\delta}$ was set at ten
percent of the natural lattice spacing while $\mu$ was varied over the
range $0 \leq \mu \leq 5$.  Since the information regarding relative
energy scales is contained within $\mu$, the specific value of
$\overline{\delta}$ does not affect the qualitative nature of the results,
so long as the displacements are small enough to maintain the topology of
the lattice.  As discussed below, however, small nonlinearities do depend
on the extent of the lattice distortions.

For all values of the control parameters, 100 different bond
configurations (i.e., realizations of the quenched disorder) were
simulated.  Calculated quantities were then averaged over the various
runs.

Two different methods of simulation were used to study the compressible
spin glass.  For the first method, suitable for studying the dynamics of
the model, states were generated via single-spinflip Monte Carlo steps,
with transition probabilities dependent upon the difference in energy
between the two spin states.  These energies employed the full Hamiltonian
of Eq.~(\ref{eqn_csg_ham}), including the components that depend on the
particle positions.  For purposes of determining transition probabilities,
the spins were considered to flip in place, i.e., without any particle
motion.  The lattice was then relaxed for the new spin configuration.  The
full simulation algorithm is as follows:  The system is started in a
random spin configuration with the particles located at the positions
which minimize the total energy.  From a given spin configuration, a
particle is chosen at random. This particle is given a chance to flip, in
place, from the state with energy $E_1$ to the state with energy $E_2$.
If $E_2 < E_1$ the spin is flipped; otherwise the spin flips with
probability $\exp [ -(E_2 - E_1) / T ]$.  After $L^2$ randomly chosen
particles have been considered (i.e., one Monte Carlo step), the lattice
is relaxed to the minimum of the potential energy for the new spin
configuration using conjugate-gradient minimization.  System properties
are recorded for analysis, and this process is then repeated.

In order to ensure proper equilibration, I follow the algorithm prescribed
by Bhatt and Young:\cite{bhat85,bhat88}  The spin-glass susceptibility
\[
   \chi_\mathrm{sg} = \frac{1}{L^2} \sum_{i,j}
         \left\langle S_i S_j \right\rangle^2  \, ,
\]
where $\langle \cdot \rangle$ indicates a thermal (time) average, is
calculated by two different methods, each as a function of equilibration
time $t_\mathrm{equil}$.  One method uses the overlap between states of
the same system at two different times during the run, while the other
uses the overlap between states of two randomly initialized, independently
run replicas of the same bond realization.  These two computation methods
produce the same value of $\chi_\mathrm{sg}$ as $t_\mathrm{equil} \to
\infty$, but the ``two-times'' method approaches the asymptotic value from
above, while the ``two-replicas'' method approaches from below.  When the
two values are within statistical error of one another, the system is
equilibrated.  The simulations are typically run for several multiples of
the equilibration time in order to acquire data from uncorrelated portions
of the time evolution.

Another method of simulation, suitable for studying static properties such
as the energy, involves substituting a collection of pre-generated spin
states into the compressible spin-glass Hamiltonian and relaxing each to
the minimum of its total energy with respect to the particle positions.
Typically, the spin states are generated by single-spinflip Monte Carlo
simulations using the standard (incompressible) spin-glass Hamiltonian,
which takes much less time than simulations of the full Hamiltonian as
described above.  In this manner, ``typical'' states may be analyzed to
observe the effect of the compressible terms on quantities of interest;
however, these states will not occur with frequency given by the correct
Boltzmann weight, so care must be taken not to draw conclusions that would
rely on such an assumption.  For the smallest system sizes ($L = 3$, 4,
and 5), it was possible to enumerate all $2^{L^2}$ possible spin states
for a given bond configuration.

For all methods of generating spin states, the lattice was relaxed to its
minimum using the conjugate-gradient minimization technique.\cite{pres97}
Since the distortions of the lattice are kept small by the value of
$\overline{\delta}$, the potential-energy landscape is close to quadratic,
and the minimum can typically be located to reasonable numerical tolerance
within a few conjugate gradient steps.  Nevertheless, because of the
computation involved in calculating the lattice energy, this portion of
the simulation takes approximately two orders of magnitude more time than
the Monte Carlo spinflips.

\section{Results}
\label{sec_results}

Simulations of the two-dimensional, constant-volume compressible Ising
spin glass were performed for system sizes ranging from $L = 3$ to 40
using the techniques described above.  Data from these simulations are
presented and anlayzed below.

\subsection{Dynamics}

\begin{figure}
\includegraphics[width=\figwidth]{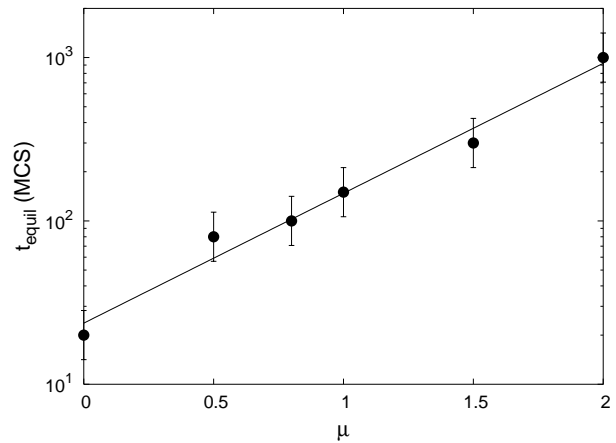}
\caption{The time to reach equilibrium, $t_\mathrm{equil}$, grows
exponentially as $\mu$ increases.  The data here are from 100 $L = 10$
systems at $T = 2.0$, and the slope of the exponential fit line is
$1.8\,\mathrm{MCS}^{-1}$.  The dramatic increase of simulation time makes
straightforward simulation of the dynamics difficult for large values of
the coupling.}
\label{fig_equil_time}
\end{figure}

The time required for the system to reach thermal equilibrium is an easily
accessible measure of the timescale for the system dynamics.  For each
value of $\mu$, a different equilibration time is required, and
Fig.~\ref{fig_equil_time} shows the dependence of the equilibration time,
$t_\mathrm{equil}$, on $\mu$ for the $L = 10$ systems at a relatively high
temperature, $T = 2.0$.  As the fit line on the semilog plot demonstrates,
the growth of the equilibration time, in Monte Carlo steps (MCS), is
exponential in $\mu$; the slope of the exponential fit is
$1.8\,\mathrm{MCS}^{-1}$.  The rapid growth of the equilibration time as
the coupling is increased can be viewed as a growth of energy barriers
between states that were previously similar in energy.  The movement of
particles ``locks in'' the current spin configuration, increasing the
timescale for single-spinflip transitions.

The growth of the equilibration time as a function of the coupling is in
addition to the usual dramatic growth of dynamic timescales as the
temperature is lowered (see, for example, Fig.~3 of
Ref.~\onlinecite{ogie85b}).  Since the number of simulation steps required
increases exponentially with $\mu$ and the computation time per step
increases in a manner proportional to the number of spins, direct
simulations of the system dynamics at temperatures approaching the
transition become prohibitive for large values of the coupling.

\subsection{Energy Analysis}

In analyzing the results of the simulations, the various components of the
total energy may be computed independently for a given spin configuration.
Of particular interest is the first term in Eq.~(\ref{eqn_csg_ham}).  This
component represents the contribution due solely to spin-spin interactions
and is denoted $E_0$.  It is equivalent to the energy of that spin
configuration on an undistorted lattice in the absence of any spin-lattice
coupling.

As shown in Fig.~1 of Ref.~\onlinecite{mars06}, the effect of the coupling
is to shift the states of the system downward in energy.  When $\mu = 0$,
the energy levels are $\delta$-functions separated by constant gaps of
$4J$, the smallest energy difference between states of the incompressible
$\pm J$ model.  As $\mu$ is increased from zero, each energy level
(identified by $E_0$) shifts downward in energy and broadens into a
Gaussian-shaped band.

\begin{figure}
\includegraphics[width=\figwidth]{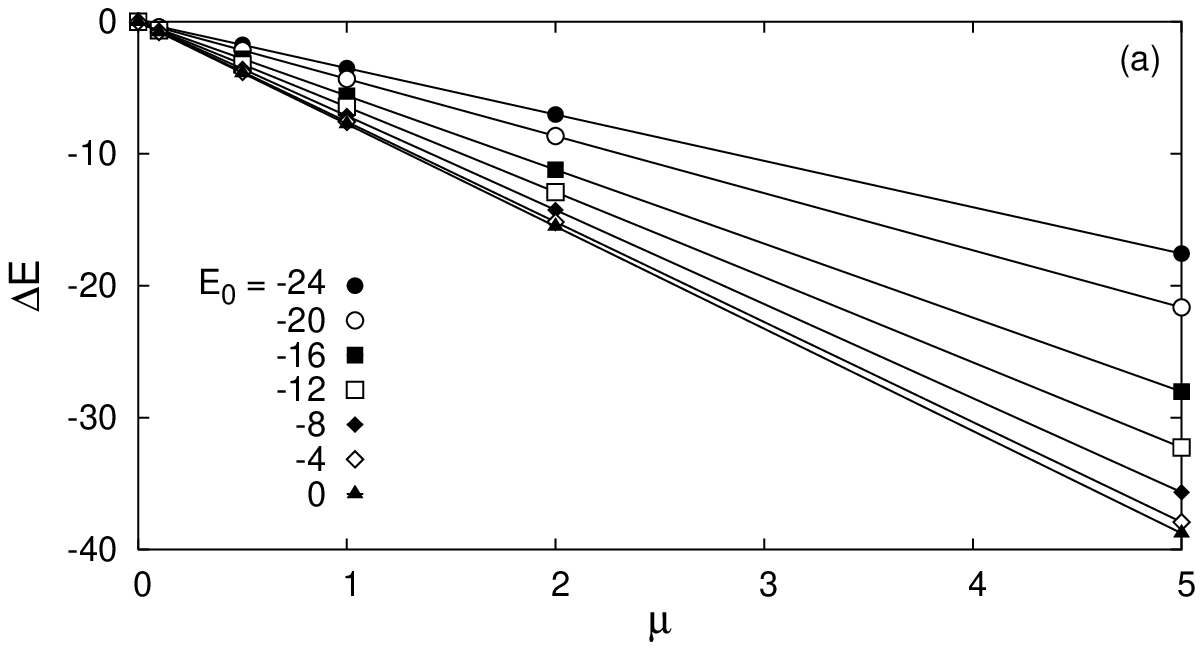}
\includegraphics[width=\figwidth]{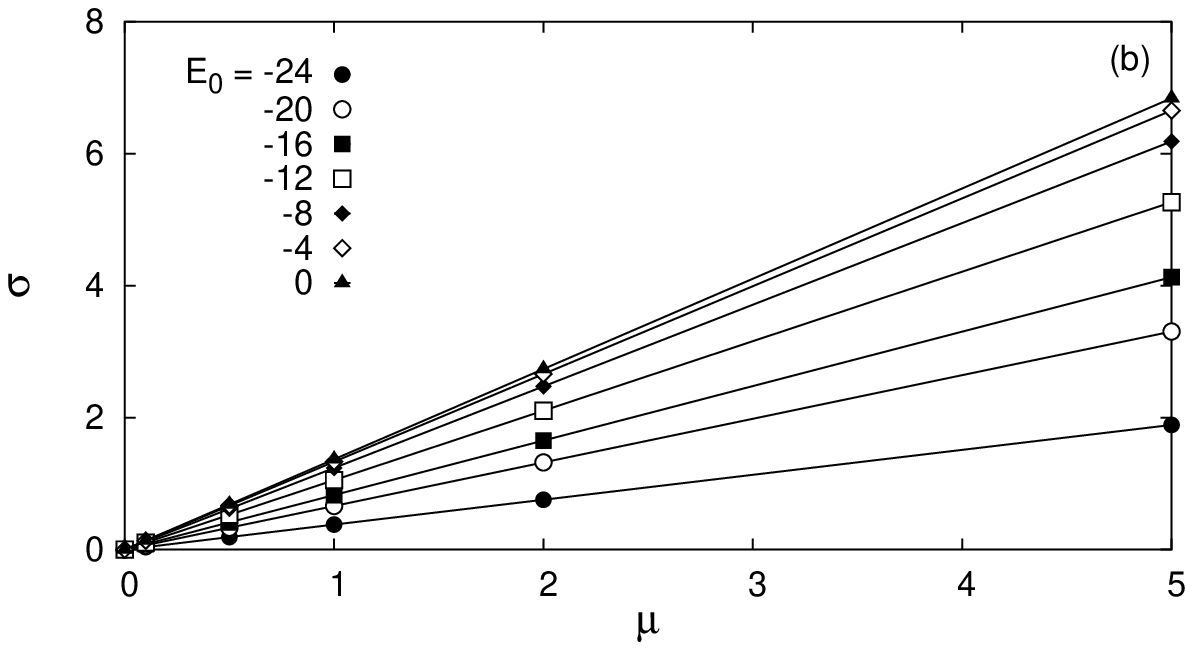}
\caption{The effect of the coupling on the energy for a single fully
enumerated $L = 4$ system.  (a)~The average shift in energy, $\Delta \!
E$, is plotted as a function of the coupling $\mu$.  For each band of
states, from $E_0 = -24$ (the ground-state energy for this particular
system) to $E_0 = 0$, the energy shifts by an amount proportional to the
coupling.  (b)~The width of each band, $\sigma$, is also proportional to
$\mu$.}
\label{fig_mu_dependence}
\end{figure}

For all of the states with a given value of $E_0$, the distribution of
energies is characterized by two values:  the average shift in energy,
$\Delta \! E(E_0,\mu) \equiv \langle E(E_0,\mu) \rangle - E_0$, and the
width $\sigma(E_0)$ as given by the standard deviation of the
distribution.  Both $\Delta \! E$ and $\sigma$ are linearly proportional
to $\mu$, as shown in Fig.~\ref{fig_mu_dependence}.  The data in that
figure were obtained from a single $L = 4$ system using complete
enumeration of all spin configurations; each line represents the data for
a value of $E_0$ ranging from the ground-state energy for this specific
system, $E_0 = -24$, to $E_0 = 0$, where there are equal numbers of
satisfied and unsatisfied bonds.

The proportional dependences of both the energy shift and the width on
$\mu$ are due to the fact that each spin state individually shifts by an
amount exactly proportional to the coupling.  When minimizing the
potential energy of the lattice for a given spin configuration, the
positions of the particles are determined by the value of
$\overline{\delta}$; the value of $\mu$ then multiplies the result to
determine the total energy in the distortions.  Due to this fact, it is
possible to characterize changes to the energy of the system at any
convenient value of $\mu$ and then scale the obtained quantites by the
coupling.

\begin{figure}
\includegraphics[width=\figwidth]{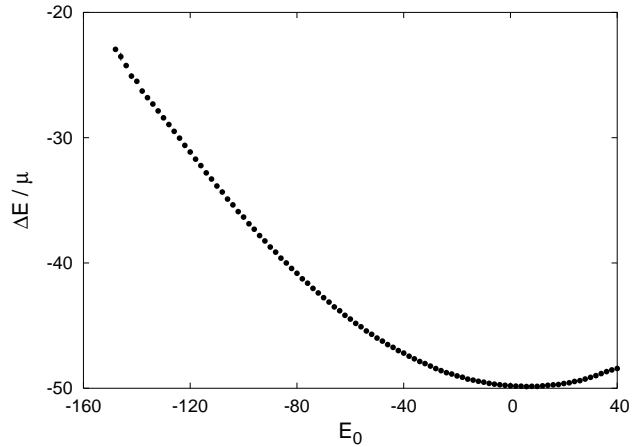}
\caption{Dependence of the energy shift on the spin-spin energy $E_0$.
These data are averaged over 100 $L = 10$ systems run at a variety of
temperatures.  The parabolic shape of this curve results from the fact
that configurations with roughly equal numbers of short (satisfied) and
long (unsatisfied) bonds can distort more effectively than those with many
bonds of the same length.}
\label{fig_energy_shift}
\end{figure}

The lines shown in Fig.~\ref{fig_mu_dependence} have different slopes,
indicating that the various bands shift and broaden at different rates as
$\mu$ increases.  The states with higher $E_0$ move downward in energy
more rapidly than lower-$E_0$ states.  Data for the shift in average
energy from the uncoupled value, scaled by $\mu$, are plotted as a
function of original energy level, $E_0$, in Fig.~\ref{fig_energy_shift};
100 systems with $L = 10$ and $\mu = 0.1$ were run at a sequence of
temperatures and averaged to produce this plot.  In practice, the states
with positive $E_0$ are difficult to populate at finite temperature due to
the exponential suppression of the Boltzmann factor.

The parabolic form of this curve can be explained by the observation that
with the volume held constant, configurations with predominantly short (or
long) bonds cannot distort as effectively as configurations with roughly
equal numbers of short and long bonds.  For even-valued system size $L$,
this curve should be symmetric about $E_0 = 0$ since there is a
relationship between spin states with alternate spins flipped: long bonds
become short bonds and vice versa, resulting in a state with $E_0$ of
equal magnitude but opposite sign that has an identical energy shift.

\begin{figure}
\includegraphics[width=\figwidth]{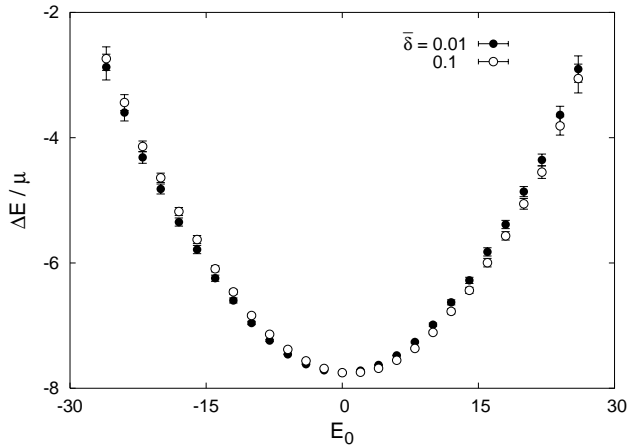}
\caption{The effect of changing the distortion parameter
$\overline{\delta}$, as defined in Eq.~(\ref{eqn_delta_def}), is shown for
the fully enumerated systems with $L = 4$ and $\mu = 0.1$.  The parabolic
form of the data is unchanged; however, the smaller value of the typical
distortion size results in a curve that is slightly more symmetric about
$E_0 = 0$.  100 systems were averaged to produce these data.} 
\label{fig_distortion}
\end{figure}

The lack of exact symmetry about $E_0 = 0$ is due to small
nonlinearities resulting from non-zero $\overline{\delta}$.
Figure~\ref{fig_distortion} shows data for the typical value of
$\overline{\delta} = 0.1$ along with a sample of data in which
$\overline{\delta}$ was set to 0.01.  The results are qualitatively
similar, though the smaller-distortion curve is more symmetric.

\subsection{Size Dependence}

\begin{figure}
\includegraphics[width=\figwidth]{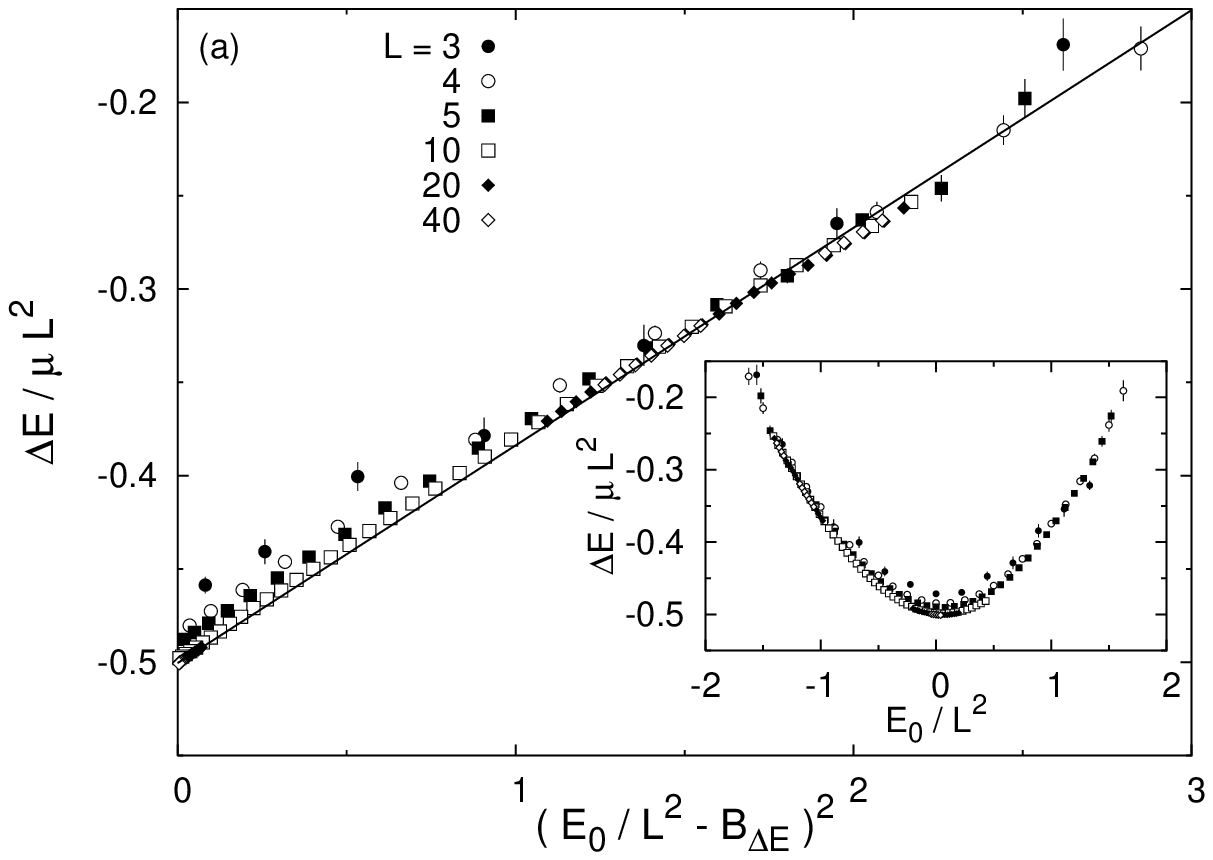}
\includegraphics[width=\figwidth]{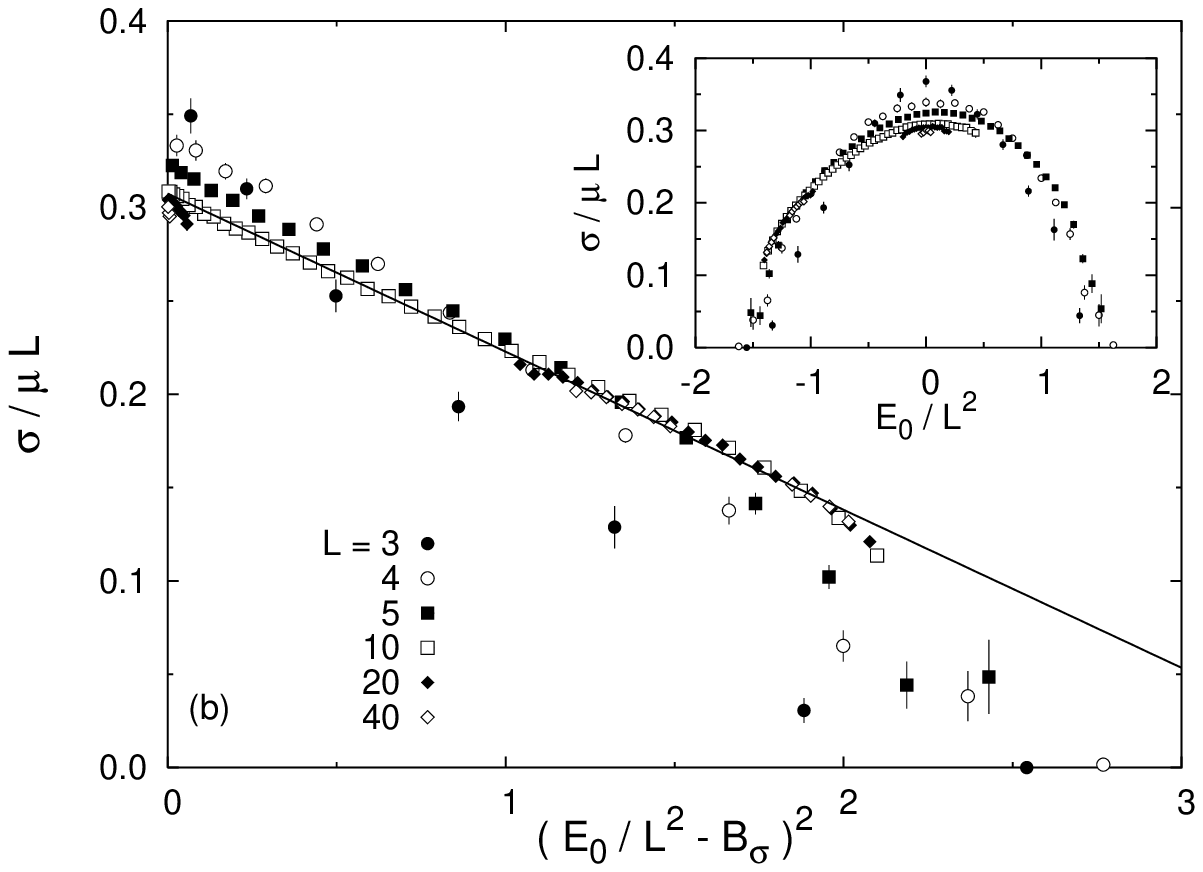}
\caption{System-size scaling.  (a)~The slope of the average energy shift,
as a function of $E_0$, with both axes scaled by $L^2$.  As $L$ increases,
the data for different system sizes collapse onto a common curve that is
quadratic in $E_0$.  The main panel shows the data plotted and fit
according to the form of Eq.~(\ref{eqn_dE_para_form}), while the inset
shows the scaled data directly.  (b)~The data for the spread in energy as
a function of $E_0$ can also be made to approach a common parabolic curve;
however, while the $E_0$-axis is again scaled by $L^2$, the width axis is
scaled by the linear size only.}
\label{fig_size_scaling}
\end{figure}

To study the size dependence of the energy, data for $\Delta \! E$ and
$\sigma$ was collected for system sizes from $L = 3$ to 40.  For all
system sizes, the shift in the average energy displays the parabolic shape
shown in Figs.~\ref{fig_energy_shift} and~\ref{fig_distortion}, and the
similarity in form suggests that the curves may be made to collapse.
Fig.~\ref{fig_size_scaling}(a) contains the results for $\Delta \! E /
\mu$ for the full range of system sizes simulated.  100 systems were
averaged at each system size; in the figure, data points are only
displayed for values of $E_0$ where at least 20 systems were represented.
The $L = 3$, 4, and~5 systems were fully enumerated, while the larger
systems were run at a series of temperatures to obtain data over a range
of values of $E_0$.  As the inset in that figure demonstrates, when both
axes are scaled by $L^2$, the data for the various system sizes approach a
constant curve as $L$ increases.  While there are finite-size effects in
the smallest systems, the data for $L \ge 10$ collapse quite well.

The quadratic form of the scaled data for $\Delta \! E$ is expressed as
\begin{equation}
   \frac{\Delta \! E}{\mu L^2} =
      A_{\Delta \! E} \times \left( \frac{E_0}{L^2} -
      B_{\Delta \! E} \right)^2 +
      C_{\Delta \! E}  \, .
\label{eqn_dE_para_form}
\end{equation}
The locations of the minima for each system size were averaged to
determine the global horizontal offset: $B_{\Delta \! E} = 0.063 \pm
0.001$.  With $B_{\Delta \! E}$ determined, the $L = 40$ data were then
fit to the parabolic form above, with $A_{\Delta \! E} = 0.1166 \pm
0.0007$ and $C_{\Delta \! E} = -0.5004 \pm 0.0008$.  The main panel of
Fig.~\ref{fig_size_scaling}(a) shows the data and fit plotted in a manner
that makes the collapse to the form of Eq.~(\ref{eqn_dE_para_form})
apparent.

Data for the width of each band also demonstrate a quadratic function of
$E_0$, as shown in Fig.~\ref{fig_size_scaling}(b).  As with the energy
shift, the $\sigma$ data for the various system sizes can be scaled to lie
on a common curve; however, while the $E_0$ axis is again scaled by the
system size $L^2$, the width axis is only scaled by the linear size of the
system, $L$.

The scaled data for the width are described by the form
\begin{equation}
   \frac{\sigma}{\mu L} =
      A_\sigma \times \left( \frac{E_0}{L^2} -
      B_\sigma \right)^2 +
      C_\sigma  \, .
\label{eqn_sigma_para_form}
\end{equation}
As with $\Delta \! E$, data from all sizes were used to obtain $B_\sigma
= 0.039 \pm 0.004$.  The $L = 40$ data were then fit to
Eq.~(\ref{eqn_sigma_para_form}), resulting in $A_\sigma = -0.085 \pm
0.002$ and $C_\sigma = 0.308 \pm 0.003$.  Figure~\ref{fig_size_scaling}(b)
shows the data and fit line.

The scaling behavior of $\sigma$ implies an interesting side effect of the
introduction of compressibility.  Since $E_0$ is proportional to $L^2$,
the total number of spins, the right-hand side of
Eq.~(\ref{eqn_sigma_para_form}) is independent of $L$, and thus $\sigma
\sim \mu L$.  Neighboring energy bands will overlap to a large degree when
the width of the bands is comparable to the spacing between them, i.e.,
when $\mu L \sim 4$.  As $L \to \infty$, an infinitesimal value of the
coupling will satisfy this condition, rendering the previously discrete
energy spectrum continuous.

\begin{figure}
\includegraphics[width=\figwidth]{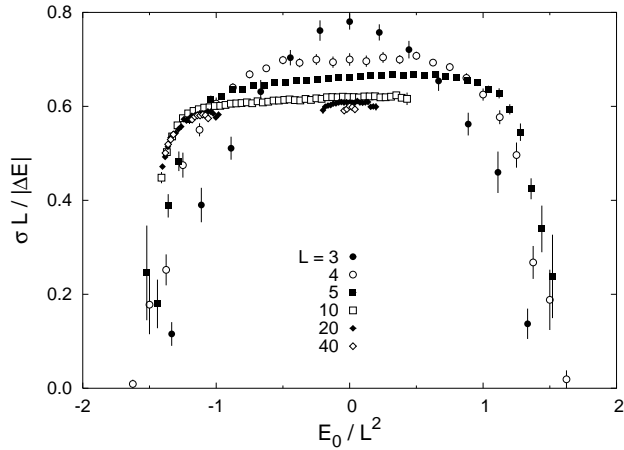}
\caption{The ratio of the width to the magnitude of the energy shift as a
function of $E_0$ for different system sizes; the axes have been scaled by
$L$ and $L^2$, respectively, to demonstrate an approach to constant
behavior as the system size increases.  Plotted this way, the scaled ratio
is less than a constant of order unity.  Thus, $\sigma$ becomes negligible
compared to $\Delta \! E$ for large $L$.}
\label{fig_width_shift_ratio}
\end{figure}

While the forms of $\Delta \! E$ and $\sigma$ are similar, it is not
immediately apparent that the two quantities are directly related.  In
fact, over a large range of $E_0$, the spread in energy is proportional to
the energy shift, as shown in Fig.~\ref{fig_width_shift_ratio}, where the
data displayed in Fig.~\ref{fig_size_scaling} are plotted as a ratio of
$\sigma \times L$ to the absolute value of $\Delta \! E$ versus $E_0 /
L^2$.  Again, the data from different system sizes were made to collapse
by appropriate scaling of the axes.

It is apparent from Fig.~\ref{fig_width_shift_ratio} that the magnitude of
the scaled ratio is less than a constant, $\rho$, with $\rho \approx 0.6$
for large values of $L$.  The relationship between $\sigma$ and $\Delta \!
E$ can be expressed as
\begin{equation}
   \sigma \lesssim \frac{\rho}{L} \left| \Delta \! E \right|  \, .
\label{eq_width_shift_ratio}
\end{equation}
Thus, as the system size increases, the width of a band of states becomes
negligible compared to the magnitude of the shift in energy from its
uncoupled value.

\section{Simplified Model}
\label{sec_simplified}

The form of $\Delta \! E$, as demonstrated in
Figs.~\ref{fig_energy_shift}, \ref{fig_distortion},
and~\ref{fig_size_scaling}(a) and expressed by
Eq.~(\ref{eqn_dE_para_form}), is a quadratic function of $E_0$.  In
addition, the spread in the energy becomes negligible compared to the
energy shift for large system sizes, as Eq.~(\ref{eq_width_shift_ratio})
demonstrates.  These observations motivate\cite{mars06} an approximate
Hamiltonian for the compressible spin glass:
\begin{equation}
   \mathcal{H}_\mathrm{approx} =
      - \sum_{\langle i,j \rangle} J_{ij} S_i S_j +
      \frac{\nu}{L^2} \left( \sum_{\langle i,j \rangle}
            J_{ij} S_i S_j \right)^2  \, .
\label{eqn_approx_ham}
\end{equation}
The sum that appears in both terms is performed as described for the
original Hamiltonian of Eq.~(\ref{eqn_csg_ham}), and numerical
factors---such as $A_{\Delta \! E}$ from Eq.~(\ref{eqn_dE_para_form})---
have been absorbed into the coupling constant so that $\nu \approx 0.12
\mu$.  Again, the first term represents the energy due to spin-spin
interactions, denoted $E_0$.  The second term, which contains the coupling
between the spins and the lattice, can be viewed as a combination of the
final two terms of Eq.~(\ref{eqn_csg_ham}) with a typical distortion of
the order of $\overline{\delta}$ as defined in Eq.~(\ref{eqn_delta_def}).

\begin{figure}
\includegraphics[width=\figwidth]{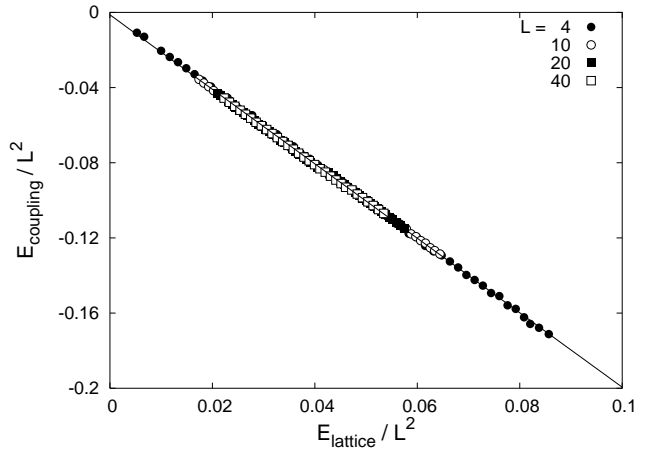}
\caption{A selection of data from various system sizes shows the
relationship between the two position-dependent components of the total
energy given by Eq.~(\ref{eqn_csg_ham}); here, the second term
(representing the energy due to the coupling, $E_\mathrm{coupling}$) is
plotted versus the third term (the lattice energy, $E_\mathrm{lattice}$)
for individual spin configurations.  Both axes are scaled by $L^2$ to
bring the data from different system sizes into a common range.  As the
coupling energy is proportional to the lattice energy, these two terms may
be combined into a single term that describes the energy shift due to
particle motions in the presence of the coupling.}
\label{fig_energy_ratio}
\end{figure}

That the two position-dependent components of the total energy may be
combined in this way is shown explicitly in Fig.~\ref{fig_energy_ratio},
where the coupling energy is plotted against the lattice energy for data
obtained in the previously described simulations.  Both axes are scaled by
$L^2$ to bring the points from different system sizes into a common range,
and the solid line is a linear fit to all of the data.  The coupling
energy is proportional to the lattice energy, and since the two terms are
related in a straightforward manner, their net effected may be represented
by a single term in the simplified Hamiltonian.

I note some features of the approximate model.  First, the system size
must be included explicitly in order to preserve the observed scaling
behavior.  Second, rather than being constructed from a combination of
parameters, the coupling constant $\nu$ is directly present and controls
the strength of the compressibility term.  Finally, and most importantly,
this Hamiltonian contains only spin degrees of freedom; the positional
variables are absent, and the degrees of freedom associated with them have
been absorbed into the second term of Eq.~(\ref{eqn_approx_ham}).  Thus,
the simplified model can be viewed as a mean-field version of the original
Hamiltonian, where the energy due to local distortions has been replaced
by an energy contribution that is typical for states with equivalent spin
energy.  As a practical matter, this feature also means that simulations
and analytical work may be performed on the model using techniques
identical to those used in standard (incompressible) spin-glass studies.

\subsection{Simulation Results}

To simulate the model described in the simplified Hamiltonian,
single-spinflip Monte Carlo simulations were performed at various values
of the coupling as parametrized by $\nu$.  As in standard Monte Carlo
spin-glass simulations, for each bond realization a sequence of spin
states was generated at a fixed temperature $T$, with transition
probabilities between states based on the difference in energy as
calculated from Eq.~(\ref{eqn_approx_ham}).

\begin{figure}
\includegraphics[width=\figwidth]{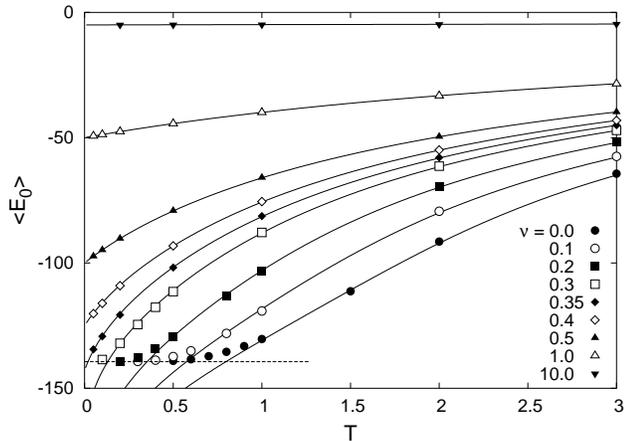}
\caption{Results from Monte Carlo simulations of the approximate model of
Eq.~(\ref{eqn_approx_ham}) on 100 systems with $L=10$.  The average value
of $E_0$ is plotted as a function of $T$ for various values of $\nu$.
Below a critical value, $\nu^* \approx 0.35$, the energy approaches that
of the ground state (dashed line) as $T \to 0$; above $\nu^*$, the energy
limit is predicted by Eq.~(\ref{eqn_energy_min}).  The solid lines are not
fits; rather, they represent predictions obtained by minimizing the free
energy using the entropy form of Eq.~(\ref{eqn_entropy_fit}).  This figure
is reprinted from Ref.~\onlinecite{mars06}.}
\label{fig_model_energy}
\end{figure}

To monitor which states of the compressible spin glass are favored at a
given value of $T$ and $\nu$, the thermal- and disorder-averaged values of
$E_0$, denoted $\langle E_0 \rangle$, are calculated.  Results for 100
systems with $L = 10$ are shown as data points in
Fig.~\ref{fig_model_energy}.  Solid lines in that figure represent
predictions based on the free-energy analysis described below in
Sec.~\ref{sec_free_energy}.

For each value of $\nu$, the average value of $E_0$ decreases as the
temperature is lowered (at infinite temperature, $\langle E_0 \rangle = 0$
since the $E_0 = 0$ states have the highest entropy).  For small values of
$\nu$, as $T \to 0$, the data approach the ground-state energy of the
uncoupled system, indicated by the dashed horizontal line in the figure.
For each curve as a function of $T$, the inflection point represents the
temperature at which the presence of the ground states of the uncoupled
system becomes important.  Below this temperature, those ground states
begin to be populated, halting the decrease in $\langle E_0 \rangle$.  As
the coupling is increased, the inflection point moves down in temperature,
eventually disappearing when $\nu \approx 0.35$.  For larger values of the
coupling, there is no inflection point, and the data for $\langle E_0
\rangle$ no longer approach the ground-state energy as $T \to 0$ but
rather terminate at some higher value that increases with increasing
$\nu$.

\begin{figure}
\includegraphics[width=\figwidth]{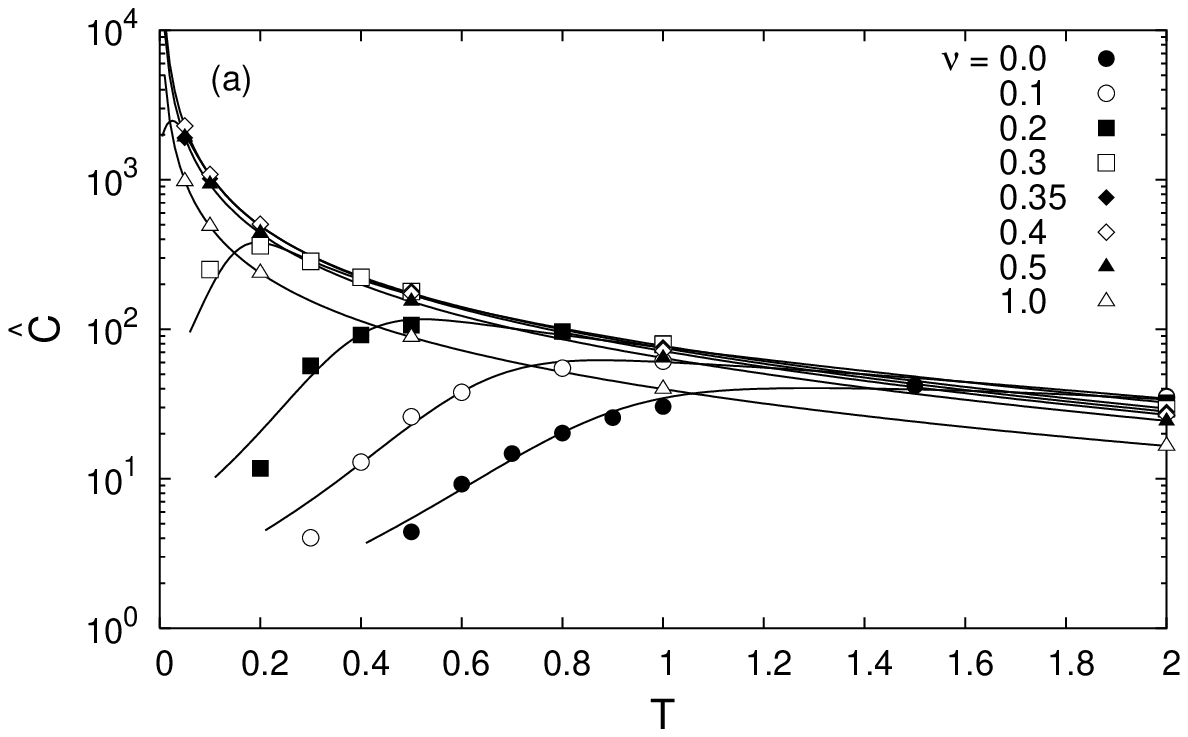}
\includegraphics[width=\figwidth]{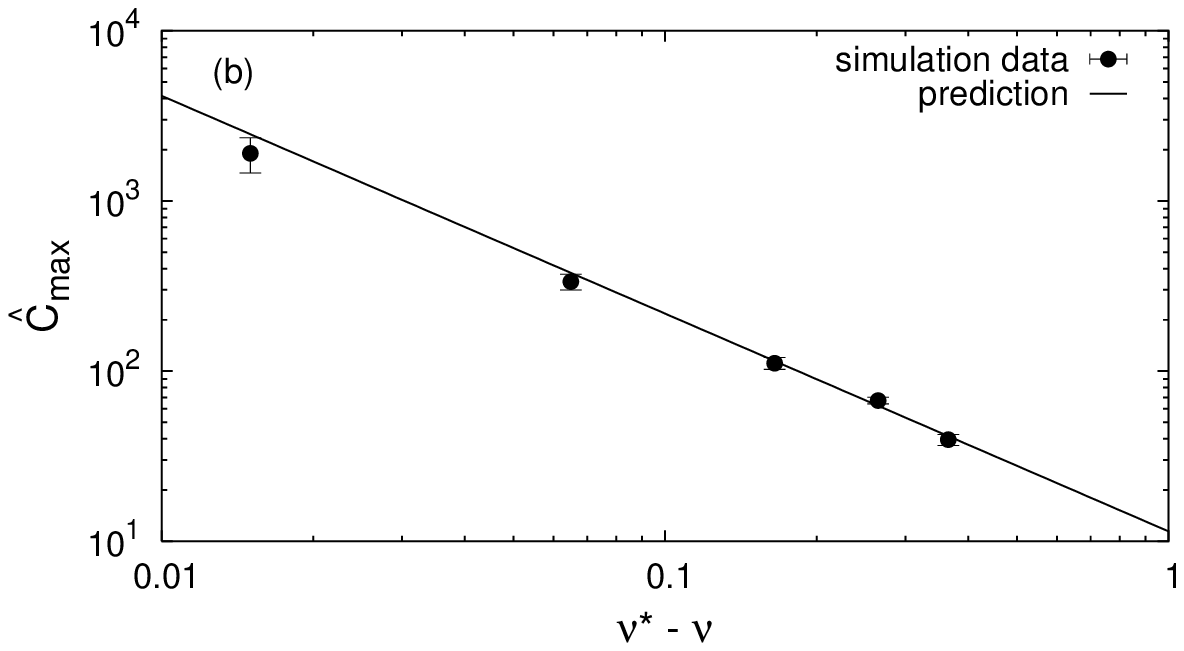}
\caption{(a)~The heat-capacity-like quantity $\widehat{C}$---defined in
Eq.~(\ref{eqn_fluctuations})---as a function of $T$ for different values
of $\nu$, calculated from simulations of the simplified model on 100
averaged systems with $L=10$.  The peak in $\widehat{C}$ shifts downward
in $T$ as $\nu$ increases.  The solid lines are predictions based on the
free-energy analysis described in Sec.~\ref{sec_free_energy}.  (b)~The
maximum value of $\widehat{C}$ demonstrates a power-law divergence as
$\nu$ approaches a critical value, $\nu^*$.  The points are obtained from
polynomial interpolation of data near the peak of each curve in panel (a).
The solid line is a fit to the peaks of the $\widehat{C}$ prediction
curves as a function of $\nu$; from the fit, $\nu^* = 0.365$, and the
power-law exponent is 1.3.}
\label{fig_model_fluctuations}
\end{figure}

The heat capacity can be calculated as the fluctuations in the energy
about its average value, divided by the square of the temperature.  For
the simplified model of the compressible spin glass, the specific heat
shows no interesting features, going smoothly to zero as $T \to 0$.
However, a similar quantity, using $E_0$ instead of the total energy, can
be calculated:
\begin{equation}
   \widehat{C} \equiv \frac{\left\langle {E_0}^2 \right\rangle -
   \left\langle E_0 \right\rangle^2}{T^2}  \, .
\label{eqn_fluctuations}
\end{equation}
Data for $\widehat{C}$ from the simulations of the simplified model are
shown as points in Fig.~\ref{fig_model_fluctuations}(a).  Solid lines in
that figure are predictions based on the free-energy analysis described
below.

At small values of $\nu$, the data and corresponding prediction for
$\widehat{C}$ are peaked.  For $\nu = 0$, i.e., the standard Ising spin
glass, this peak is interpreted as signaling the onset of critical
behavior that precedes the spin-glass transition as the temperature
continues to be lowered.\cite{ogie85b}  As the coupling approaches a
critical value, $\nu^*$, the temperature at which the peak in
$\widehat{C}$ is located moves toward zero, and the height of the peak,
$\widehat{C}_\mathrm{max}$, diverges.
Figure~\ref{fig_model_fluctuations}(b) shows that this divergence is a
power law:
\[
   \widehat{C}_\mathrm{max} = A \left( \nu^* - \nu \right)^{-p}  \, .
\]
The points in that figure are the locations of the maxima, as obtained
from parabolic fits to data near the peak of each curve as a function of
$T$, while the solid line is a power-law fit to the peaks of the predicted
curves with $\nu^* = 0.365$ and a power-law exponent of 1.3.  The reasons
for the divergence are discussed below.

\subsection{Analytic Results}

For the approximate Hamiltonian of Eq.~(\ref{eqn_approx_ham}), there is a
one-to-one correspondence between the spin energy $E_0$ (calculated as
before) and the total energy.  To understand the results of the simulation
as presented in Fig.~\ref{fig_model_energy}, it is useful to analyze the
expected value of the energy for various values of the control parameters.
In simplified notation, Eq.~(\ref{eqn_approx_ham}) can be
written\cite{mars06}
\begin{equation}
   E = E_0 + \frac{\nu}{L^2} {E_0}^2 \, .
\label{eqn_model_energy}
\end{equation}
Taking the derivative with respect to $E_0$ allows one to calculate the
value of the spin energy, $E_{0,\mathrm{min}}$, that minimizes the total
energy as function of $\nu$:
\begin{equation}
   E_{0,\mathrm{min}} = \frac{-L^2}{2 \nu} \, .
\label{eqn_energy_min}
\end{equation}
This represents the expected value of the spin energy as $T \to 0$;
however, the calculated value is not necessarily realizable, since $E_0$
must be greater than the ground-state spin energy $E_{0,\mathrm{gnd}}$.
For small values of $\nu$, the calculated value of $E_{0,\mathrm{min}}$
lies in the non-physical region below $E_{0,\mathrm{gnd}}$, and thus the
minimum spin energy is, of course, equal to the ground-state spin energy.
This explains the form of the small-$\nu$ data in
Fig.~\ref{fig_model_energy}, where the average spin energy decreases with
temperature but approaches $E_{0,\mathrm{gnd}}$ asymptotically as $T \to
0$.

There is a value of $\nu$ at which the minimum $E_0$ becomes equal to the
ground-state energy of the uncoupled system, $E_{0,\mathrm{gnd}}$:
\begin{equation}
   \nu^* \equiv \frac{-L^2}{2 E_{0,\mathrm{gnd}}} \, .
\label{eqn_nu_crit}
\end{equation}
For the two-dimensional spin-glass, the ground-state energy per spin is
$-1.4$,\cite{wang88} and thus $\nu^* = 0.36$.  This value is the same as
the value of $\nu$ at which the simulation data, as shown in
Figs.~\ref{fig_model_energy} and~\ref{fig_model_fluctuations}, display a
change in behavior.

At $\nu = \nu^*$, the nature of the energy spectrum is dramatically
altered:  As the coupling increases from zero, higher-$E_0$ states shift
downward in energy more rapidly than lower-$E_0$ states, and thus the
difference in total energy between neighboring $E_0$ levels becomes
smaller.  The value $\nu^*$ represents the coupling at which the ground
state and first excited state of the uncoupled system have the same total
energy.  Above this value, states with $E_0 = E_{0,\mathrm{gnd}}$ no
longer have the lowest total energy, and as $\nu$ increases still more,
increasingly higher $E_0$-levels are associated with the ground states of
the compressible system.  The data in Fig.~\ref{fig_model_energy} display
this feature, as the zero-temperature value of $\langle E_0 \rangle$
increases with the coupling for $\nu > \nu^*$.  The divergence in
$\widehat{C}$, as shown in Fig.~\ref{fig_model_fluctuations}, is also a
consequence of this change in the energy spectrum.  As $\nu \to \nu^*$,
the difference in total energy between levels near $\langle E_0 \rangle$
goes to zero, and thus the fluctuations in $E_0$ no longer vanish as $T
\to 0$.

\subsection{Free Energy Analysis}
\label{sec_free_energy}

In order to predict which states are preferred as a function of $\nu$ and
$T$, the free energy must be minimized, and for this a functional form for
the entropy is needed.  The probability of generating a state of given
energy is proportional to the Boltzmann-weighted density of states:  $P(E)
\propto \Omega(E) \exp(-E/T)$; from this the entropy $S(E)$ is derived as
$\log \Omega$.  Since the density of states is a function of the uncoupled
energy, i.e., $\Omega = \Omega \left( E(E_0) \right)$, it may be calculated
for any value of the coupling.  For a given $\nu$ and $T$, simulations
will produce a limited range of energies that will be populated with
statistical significance; data is therefore acquired at different
couplings and temperatures to produce overlapping regions of data that may
be combined.  Since for each run the proportionality between the generated
probabilities and the density of states is unknown, it is more convenient
to generate the derivative of the entropy:
\begin{equation}
   \frac{dS}{dE_0} = \frac{d}{dE_0} \left[ \log P\left( E_0 \right) +
         E\left( E_0 \right) / T \right] \, .
\label{eqn_entropy_deriv}
\end{equation}

\begin{figure}
\includegraphics[width=\figwidth]{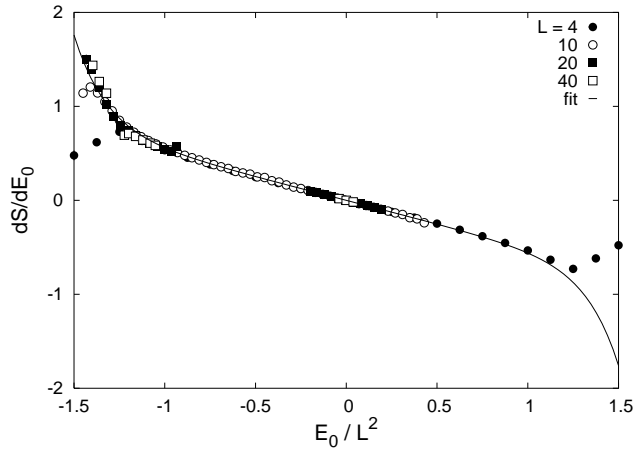}
\caption{Data for the derivative of the entropy, averaged over 100
systems each at sizes from $L = 4$ to 40.  When plotted as a function of
$E_0$ per spin, the data lie on a common curve that is well fit by the
functional form given in Eq.~(\ref{eqn_entropy_deriv_fit}).}
\label{fig_entropy_deriv}
\end{figure}

Data for the derivative of the entropy is shown in
Fig.~\ref{fig_entropy_deriv}.  For each system size, 100 individual bond
configurations were run at a variety of temperatures and averaged.  By
plotting $dS/dE_0$ versus $E_0$ per spin, the data from different system
sizes are made to lie on a single curve.  This curve is linear over a
large region passing through $E_0 = 0$, and the deviations from linearity
are exponential.  As demonstrated by the solid line in
Fig.~\ref{fig_entropy_deriv}, the overall curve is well fit by the
functional form
\begin{equation}
   \frac{dS}{dE_0} = -c_1 \frac{E_0}{L^2} -
         c_2 \sinh \left( c_3 \frac{E_0}{L^2} \right) \, ,
\label{eqn_entropy_deriv_fit}
\end{equation}
where $c_1 = 0.5$, $c_2 = 4.5 \times 10^{-4}$, and $c_3 = 5.6$.  The
entropy is thus of the form
\begin{equation}
   S \left( E_0 \right) = S_0 - S_1 {E_0}^2 -
         S_2 \cosh \left( S_3 E_0 \right)  \, ,
\label{eqn_entropy_fit}
\end{equation}
with $S_1 = 0.25 / L^2$, $S_2 = 8.1 \times 10^{-5} L^2$, and $S_3 =
5.6 / L^2$.

With the entropy given by Eq.~(\ref{eqn_entropy_fit}) and the energy given
by Eq.~(\ref{eqn_model_energy}), the free energy, $F = E(E_0) - T S(E_0)$,
may be minimized with respect to $E_0$.  The spin energy of the system in
the thermodynamic limit is thus given by the solution to the equation
\[
   1 + \frac{2 \nu}{L^2} E_0 + 2 T S_1 E_0 +
         T S_2 S_3 \sinh \left( S_3 E_0 \right) = 0\, .
\]
While this equation cannot be solved analytically, it is possible to
obtain a numerical solution as a function of $\nu$ and $T$.  Such results
are plotted as solid lines in Fig.~\ref{fig_model_energy}, where the
values of $\langle E_0 \rangle$ as predicted from the free-energy
calculation are in excellent agreement with those obtained from
simulations of the simplified model.  The predictions tend to diverge from
the data at low temperatures for small values of the coupling; this is the
regime in which the lowest-energy states are heavily populated and the
functional form for the entropy---which contains no low-energy
cutoff---ceases to be a good description of any actual system.

The free energy described above was also used to predict values for the
heat-capacity-like quantity $\widehat{C}$, as defined in
Eq.~(\ref{eqn_fluctuations}).  These predictions are shown for various
values of the coupling as the solid lines in
Fig.~\ref{fig_model_fluctuations}(a); the agreement with data from the
simulations is very good except at low temperatures for small $\nu$, where
the applicability of this form for free energy is expected to break down.
Near the peaks in $\widehat{C}$, however, the predictions are still in
reasonable agreement with the simulation results.  The peaks of the
predicted curves were thus used to generate a prediction for the
divergence of $\widehat{C}_\mathrm{max}$.  The power-law fit to these
peaks is shown as the solid line in Fig.~\ref{fig_model_fluctuations}(b),
where the power-law exponent is 1.3 and $\nu^* = 0.365$, consistent with
the value of 0.36 obtained via Eq.~(\ref{eqn_nu_crit}).

\subsection{Phase Diagram}

\begin{figure}
\includegraphics[width=\figwidth]{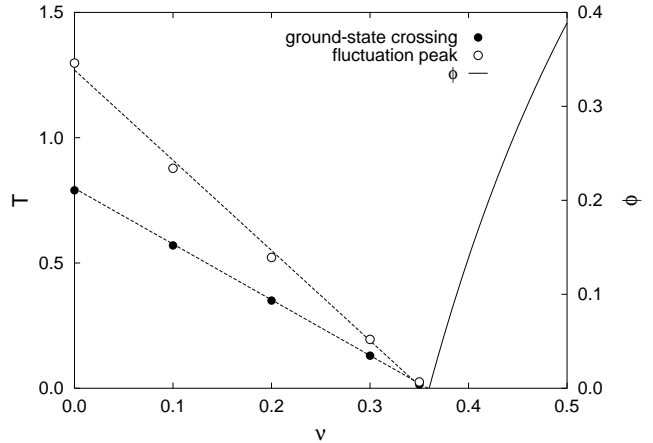}
\caption{Phase diagram in the $\nu-T$ plane showing the temperature at
which the predicted energy crosses the ground-state energy and the
temperature at which the peak in $\widehat{C}$ occurs.  Dashed lines
indicate linear fits to these data; the two lines terminate at $\nu^*$.
The region beneath the lines represents the critical regime, signaling the
onset of the spin-glass phase.  At the same value $\nu^*$, the order
parameter $\phi$---defined in Eq.~(\ref{eqn_order_param})---increases
linearly from zero, indicating the suppression of critical behavior.}
\label{fig_phase_diagram}
\end{figure}

It is possible to map out various quantities as a function of $\nu$ and
$T$.  Consider the temperature at which the predicted value of $\langle
E_0 \rangle$ crosses the ground-state energy of the uncoupled system (see
Fig.~\ref{fig_model_energy}).  This represents the breakdown of the
prediction due to the lack of a consistent analytic form for the entropy
near the ground state.  Near this temperature, there is an inflection
point in the curve of the average $E_0$ as the presence of the ground
state becomes important and the curve begins to flatten.  Also of interest
is the location of the peak in $\widehat{C}$ with respect to $T$, which
indicates the onset of critical spin-glass behavior.  Both of these
quantities, calculated from the predictions described in
Sec.~\ref{sec_free_energy}, are plotted in a $\nu-T$ phase diagram in
Fig.~\ref{fig_phase_diagram}, with linear fits to the points.

The lines in the $\nu-T$ plane mark an approximate phase boundary between
the normal (paramagnetic) phase and the critical regime that signals the
onset of spin-glass behavior.  The fit lines for the ground-state crossing
and the peak in $\widehat{C}$ terminate at $\nu = 0.359$ and 0.353,
respectively, consistent with the value of $\nu^*$ predicted from
Eq.~(\ref{eqn_nu_crit}).  Commensurate with the termination of these lines
at $\nu^*$ is the growth from zero of an order parameter characterized by
the difference between the minimum value of $E_0$ and the ground-state
energy of the uncoupled system:
\begin{equation}
   \phi = E_{0,\mathrm{min}} - E_{0,\mathrm{gnd}}  \, .
\label{eqn_order_param}
\end{equation}
Using Eqs.\ (\ref{eqn_energy_min})~and~(\ref{eqn_nu_crit}), it is apparent
that $\phi$ grows as $1/\nu^* - 1/\nu$, i.e., linear just above the
critical value $\nu^*$.  Figure~\ref{fig_phase_diagram} shows this
phenomenon.

The interpretation of $\phi$ is as a measure of the inaccessibility of
low-$E_0$ states, even at low temperatures, due to the presence of
the coupling to lattice distortions.  These states, at and near the ground
state of the uncoupled system, are no longer the lowest-energy states of
the compressible spin glass, and the competition between energy and
entropy no longer exists.  Non-zero $\phi$ is thus correlated with the
suppression of critical behavior that precedes the spin-glass phase; above
$\nu^*$, the spin-glass transition cannot exist.

\section{Conclusions}
\label{sec_conclusion}

This paper has analyzed a model\cite{mars06} for a compressible Ising spin
glass that lends itself to simulations similar to those for standard spin
glasses, with additional steps to determine the positions of the spin
particles.  While exploration of the dynamics of this system has proven
difficult, it is possible to characterize the effect of the coupling to
lattice distortions on the energy of the system.  Both the shift in energy
and the width of each band of states display parabolic shapes as a
function of the uncoupled energy $E_0$, and system-size scaling
demonstrates that the width becomes negligible compared to the energy
shift as $L$ increases.

The form of the shift in energy due to the presence of compressibility
motivates a simplified model for the compressible spin glass.\cite{mars06}
This model, which depends only upon spin degrees of freedom, was simulated
using standard techniques.  In addition, analysis of the simplified model
suggests a critical value of the coupling above which the nature of the
energy levels changes dramatically, and the simulation data confirm this.
Due to the elimination of the critical regime, a spin-glass transition
cannot exist above this critical value.

The simplified model adds long-range interactions to the nearest-neighbor
behavior of the standard Edwards-Anderson model for the spin glass.  This
provides a convenient mechanism for incorporating phonon effects into
theoretical spin-glass studies, and it is possible that consideration of
these effects may help to shed light on some experimental results that
have yet to be fully explainedr.  For example, work by Bitko et
al.\cite{bitk96} demonstrated the existence of a signature for the
spin-glass transition at high frequencies, hinting that coupling to
high-frequency modes (such as phonons) may be important.

I expect the results not to change qualitatively in three dimensions.
Preliminary studies similar to those described above suggest that, as for
the two-dimensional case, the shift in energy is proportional to the
coupling, $\mu$, and to ${E_0}^2$.  Furthermore, the energy shift scales
as the volume of the system, $L^3$, while the spread in each energy band
scales as $L$.  Thus, the assumptions that led to the simplified model of
Eq.~(\ref{eqn_approx_ham}) hold even more strongly in three dimensions;
similar results for the elimination of the spin-glass transition above a
critical value of the coupling should then follow, but additional work is
required to verify this.

It is important to note that many of these results are expected to be
quite different if the constraint of constant volume is removed.  The
quadratic nature of the energy shift (as shown in
Figs.~\ref{fig_energy_shift}, \ref{fig_distortion},
and~\ref{fig_size_scaling}) depends on the fact that states at large
negative (positive) $E_0$ cannot distort effectively due to having large
numbers of satisfied (unsatisfied) bonds that tend to have similar
lengths.  A system capable of uniform compression or expansion could take
advantage of these states with extreme values of $E_0$ in an entirely
different manner than a system where the volume is constant.

%
%

\begin{acknowledgments}
I would like to thank S.\ Nagel and B.\ Chakraborty for professional
guidance and S.\ Coppersmith, G.\ Grest, S.\ Jensen, J.\ Landry, N.\
Mueggenburg, and T.\ Witten for helpful discussions.  This work was
supported by NSF DMR-0352777 and MRSEC DMR-0213745.
\end{acknowledgments}


\end{document}